\documentstyle{depart}
\begin{document}
\baselineskip22pt

\begin{centering} Standard Clocks, Orbital Precession and\\ the Cosmological Constant\\
\vspace{.15in} Andrew W. Kerr, John C. Hauck, and Bahram Mashhoon\\
\vspace{.15in} Department of Physics and Astronomy\\ University of Missouri-Columbia

Columbia, Missouri 65211, USA

\end{centering}
\vspace{.5in}
\begin{center} Abstract
\end{center} We discuss the influence of the cosmological constant on the gravitomagnetic
clock effect and the gravitational time delay of electromagnetic rays.  Moreover, we
consider the relative motion of a binary system to linear order in the cosmological
constant $\Lambda$.  The general expression for the effect of
$\Lambda$ on pericenter precession is given for arbitrary orbital eccentricity.\\


\newpage

\noindent 1.  Introduction

\vspace{.10in} Current cosmological models that take the acceleration of the universe into
account involve a cosmological constant $\Lambda$ of magnitude $\Lambda\approx H^2_0/c^2$,
where $H_0$ is the Hubble parameter. Taking $H_0\approx 70\,{\rm km\:sec^{-1}\,Mpc^{-1}}$, we
find that $\Lambda\approx 10^{-56}{\rm cm}^{-2}$.  The local effects of such a cosmological
constant are expected to be very small.  Nevertheless, great strides are being made in
observational techniques in astronomy and it appears worthwhile to investigate some of the
observational consequences of the existence of a cosmological constant.

The dependence of the gravitomagnetic clock effect on the cosmological constant is examined
in section 2.  Section 3 is devoted to the gravitational time delay and its dependence on
$\Lambda$.  The influence of the cosmological constant on the motion of a binary system is
investigated in detail in sections 4 and 5.  A bound orbit is shown to precess with a
frequency given by $\frac{1}{2}(\Lambda c^2/\omega)\sqrt{1-e^2}$, where $\omega$ and $e$ are
the orbital frequency and eccentricity, respectively.  Section 6 contains a brief discussion
of our results.

\vspace{.22in}

\noindent 2. Clock effects in the Kerr-de Sitter spacetime

\vspace{.10in} The gravitomagnetic clock effect is caused by the net spin of a gravitational
source [1-4].  In general, the rotation of a massive body introduces a difference in the co-
and counter-rotating orbital periods of test masses in motion around the central source.  We
wish to investigate this gravitomagnetic effect in the equatorial plane of the Kerr-de Sitter
spacetime to find what influence the cosmological constant would have on the clock effect.   
The Kerr-de Sitter metric in the standard Boyer-Lindquist coordinates
$x^{\mu} = (t, r,
\theta, \phi)$ is given by [5, 6]

\begin{eqnarray} ds^2=&-& \left[ 1 - \frac{2Mr}{\Sigma} - \frac{\Lambda}{3} (r^2 + a^2\; {\rm
sin}^2\theta)\right] dt^2  - 2a\left[\frac{2Mr}{\Sigma} + \frac{\Lambda}{3} (r^2 +
a^2)\right]\;{\rm sin}^2\theta dtd\phi\nonumber\cr\\ &+& \frac{\Sigma}{\Delta} dr^2 +
\frac{\Sigma}{\chi} d\theta^2 + \left[ \frac{2Mr}{\Sigma} a^2{\rm sin}^2\theta + (1+
\frac{\Lambda}{3} a^2) (r^2 + a^2)\right] \; {\rm sin}^2\theta d\phi^2\;\;,
\end{eqnarray}

\noindent where units are chosen such that $G=c=1$, unless specified otherwise, and 

\begin{equation} 
\Sigma = r^2+ a^2\cos^2\theta\;\;,\;\;\chi = 1 + \frac{\Lambda}{3} a^2\cos^2\theta\;\;,
\end{equation}

\begin{equation}
\Delta = r^2 - 2Mr + a^2 - \frac{\Lambda}{3} r^2(r^2 + a^2)\;\;.
\end{equation}

\noindent The mass of the source is $M, J = Ma$ is its angular momentum and $\Lambda$ is the
cosmological constant.  The metric (1) represents a rotating Kottler spacetime [7].  The
Kottler spacetime is also known as the Schwarzschild-de Sitter spacetime.  Useful background
information on de Sitter spacetime and the cosmological constant can be found in [8] and
[9,10], respectively.

It is interesting to note that in the Newtonian limit the physical content of equation (1)
reduces to a Newtonian gravitational potential $\Phi_N = -GMr^{-1} - \frac{1}{6}\Lambda c^2
r^2$, which satisfies the Poisson equation $\nabla^2\Phi_N = 4\pi G\rho_N$ with $\rho_N =
M\delta({\bf r})+ \rho_{\Lambda}$.  Here $\rho_{\Lambda} = -c^2\Lambda/(4\pi G)$ is the
effective negative uniform density of ``matter'' represented by the cosmological constant. 
The effective repulsive force due to this source is then given by $\frac{1}{3}\Lambda c^2{\bf
r}$ per unit mass.  On the other hand, the relativistic interpretation of the cosmological
constant in terms of vacuum energy involves a ``substance'' with equation of state
$\hat{\rho} c^2 + \hat{p} = 0$ such that the density $\hat{\rho}$ is positive, but the
pressure is negative and is given by $\hat{p} = -c^4\Lambda/(8\pi G)$.

Let us now consider circular geodesics of this spacetime.  The stability of circular geodesic
orbits in the equatorial plane of Kerr-de Sitter geometry has been investigated by Howes
[11].  It turns out that stable circular orbits exist out to a radius of
$(3M/\Lambda)^{1/3}$.  On the other hand, the inner stability limits vary depending on the
sense of orbital motion as well as the values of $a$ and $\Lambda$ [11].  For $\Lambda = 0$,
we recall that for the corotating case, the inner stability threshold varies from $r = 6M$ to
$r= M$ as $a: 0 \rightarrow M$, whereas it varies from $r= 6M$ to $r=9M$ in the
counterrotating case.  

The geodesic equation for the radial coordinate reduces to [12, 13]

\begin{equation} g_{tt, r} \left(\frac{dt}{d\phi}\right)^2 + 2g_{t\phi,
r}\left(\frac{dt}{d\phi}\right) + g_{\phi\phi, r} = 0\;\;,
\end{equation}

\noindent which has the solution

\begin{equation}
\frac{dt}{d\phi} = a \pm \omega_K^{-1}\;\;.
\end{equation}

\noindent Here $\omega_K$ is the modified Kepler frequency

\begin{equation}
\omega_K^{2} = \frac{M}{r^3} - \frac{1}{3}\Lambda\;\;,
\end{equation}

\noindent which can be interpreted as being due to a net mass of $M +
(4\pi/3)\rho_{\Lambda}r^3$, where $\rho_{\Lambda} = -\Lambda/4\pi$ is the constant Newtonian
density mentioned before.  In connection with equation (6) as well as other
modifications of Kepler's third law to include dark matter, etc., we note that the
motion of planets in the solar system can be used to study deviations from
Keplerian motion; however, for $\Lambda\approx 10^{-56}\:{\rm cm}^{-2}$ the
corresponding effects are too small to be detectable at present.  

There are three distinct clock effects that can be studied using counterrevolving circular
geodesic orbits.  To find the influence of the cosmological constant on the clock effects, we
consider these in turn.  Integration of equation (5) over $2\pi$ for corotating and
$-2\pi$ for counterrotating test particles implies that $t_{\pm} = T_K\pm 2\pi a$, where $T_K
= 2\pi/\omega_K$ is the Keplerian period.  Thus $t_+ - t_- = 4\pi a$, just as in the Kerr
case.  Consider now an (accelerated) standard clock at rest on the circle of radius $r$ in
the equatorial plane.  A standard clock measures the proper time along its world line. 
According to this clock, the difference in the time $\tau^{\prime}$ that it would take two
free counterrevolving test particles to complete the orbit is

\begin{equation}
\tau^{\prime}_{+} - \tau^{\prime}_- = 4\pi a\;\sqrt{1-\frac{2M}{r} - \frac{\Lambda}{3}(r^2 +
a^2)}\;\;,
\end{equation}

\noindent which is the observer-dependent single-clock clock effect according to the
terminology of [12].  For the cosmological constant to have any significant influence on
this clock effect, the orbital radius would have to be unreasonably large.  In fact,
$\Lambda\approx 10^{-56}\:{\rm cm}^{-2}$ has a totally negligible effect even at a planetary
radius of $r\approx 10\:{\rm AU}$, since $\Lambda r^2/3\approx 10^{-28}$.   

Consider now the proper periods of revolution $\tau_{\pm}$ of the two free standard clocks
counterrevolving on the circular orbit of radius $r$; we find from equations (1) and (5) that

\begin{equation}
\tau_{\pm} = T_K\sqrt{1-\frac{3M}{r} - \frac{\Lambda}{3}a^2\pm2 a \omega_K}\;\;.
\end{equation} 

\noindent Expanding this expression in powers of $\epsilon = a/M$, we find that 

\begin{equation}
\tau_+ - \tau_- = \frac{4\pi a}{\sqrt{1-\frac{3M}{r}}}\left[1+\frac{Ma^2(1-\Lambda
r^2)}{2r(r-3M)^2} + O(\epsilon^4)\right]\;\;.
\end{equation}

\noindent Thus the cosmological constant does not affect the observer-dependent two-clock
clock effect to first order in $a/M$.  

Imagine now two standard clocks starting at the event characterized by $(0, r, \pi/2, 0)$,
i.e. $t=0$ at $\phi=0$, and moving freely in opposite directions on the circle of radius
$r$.  Since the corotating clock is slower, their first meeting occurs at $\phi_1 = \pi -
\alpha$, where $\alpha = a\omega_K$.  Let their nth meeting point be the event characterized
by $t_n$ and $\phi_n$, where $t_n = \frac{1}{2}nT_K(1-\alpha^2)$ and $\phi_n=n\pi(1-\alpha)$
modulo $2\pi$.  The observer-independent two-clock clock effect refers to the difference in
the proper times $\tau^+_n - \tau^-_n$ of the clocks at their nth meeting, where

\begin{equation}
\tau^{\pm}_n =
\frac{1}{2}nT_K(1\mp\alpha)\sqrt{1-\frac{3M}{r}-\frac{\Lambda}{3}a^2\pm2\alpha}\;\;.
\end{equation} 

\noindent To first order in $a/M$, this clock effect is also independent of $\Lambda$ and is
given by 

\begin{equation}
\tau^+_n - \tau^-_n\approx \frac{6\pi nJ}{\sqrt{r(r-3M)}}\;\;.
\end{equation}

\noindent The diametrical line joining the points of encounter with the origin of the
circular orbit undergoes a precession in the opposite sense as the rotation of the source. 
The precession frequency is given approximately by $n\pi\alpha/\tau^+_n$, which to lowest
order coincides with the precession frequency of a fixed torque-free test gyroscope in the
equatorial plane at radius $r$.

The gravitomagnetic clock effect (9) far from the source, $3M<<r<<(3M/\Lambda)^{1/3}$, is
essentially independent of $\Lambda$, and is basically proportional only to $a=J/M$, which is
the specific angular momentum of the source.  That is, the effect has a topological character
as it is practically independent of the radius of the orbit.  It is also essentially
independent of the gravitational constant $G$; therefore, the effect can be ``large''.  For
instance, for space-borne clocks around the Earth the effect is $\sim10^{-7}$ sec; however,
the detection of the effect requires knowledge of satellite orbits to millimeter accuracy
[3].  This seems to be somewhat beyond present capabilities by about an order of magnitude;
moreover, the influence of the cosmological constant on the clock effect is too small to be
detectable in the foreseeable future.  Another aspect of the clock effect is that it takes
longer to complete a prograde circular orbit in the equatorial plane than the corresponding
retrograde orbit.  Thus free motion in the same sense as the rotation of the source is slower
than motion in the opposite sense, which is contrary to what one might expect from the
``dragging of space'' by the rotating  source.

There is a close relationship between the clock effect and circular holonomy.  In fact,
circular holonomy in the Kerr-de Sitter spacetime can be developed in close analogy with that
in the Kerr spacetime [14].  This would involve the parallel transport of a vector around a
constant-time circular $\phi$-loop in the equatorial plane of the Kerr-de Sitter spacetime. 
Let us just mention here the result that there is a band of holonomy invariance for
$r_*<r<(3M/\Lambda)^{1/3}$ given by $\tilde {n}f(r) = \tilde{m}$, where $\tilde{n}$ and
$\tilde{m}$ are positive integers and 

\begin{equation} 
f^2(r) = 1-\frac{2M}{r} - \frac{2Ma^2}{r^3} -
\frac{M^2a^2}{r^4}-\frac{1}{3}\Lambda r^2 + \frac{1}{3}\Lambda a^2\left(1 +
\frac{2M}{r}\right)
\end{equation}

\noindent to first order in the cosmological constant.  Here $r_*$ is the unique positive
root of $f(r) = 0$ for sufficiently small $\Lambda$.  

\vspace{.25in}

\noindent 3. Time delay

\vspace{.10in} Let us consider the time delay in the propagation of electromagnetic rays in
the Kerr-de Sitter spacetime.  The influence of the cosmological constant on the local
propagation effects is expected to be very small; therefore, it is sufficient to consider the
linearized Kerr-de Sitter metric in {\it isotropic} coordinates on the background Minkowski
spacetime.  In this case, the time that it takes for the rays to travel from the point
$P_1:(t,\mbox{\boldmath$\rho_1$})$ to the point $P_2: (t, \mbox{\boldmath$\rho_2$})$ can be
written as 

\begin{equation} t_2-t_1 = |\mbox{\boldmath$\rho_2$} - \mbox{\boldmath$\rho_1$}| +
\Delta_{GE} + \Delta_{GM} +
\Delta_{\Lambda}\;\;,
\end{equation}

\noindent where $\mbox{\boldmath$\rho$}$ is the position vector in the background Minkowski
spacetime and $\Delta_{GE}$ and $\Delta_{GM}$ respectively represent the standard
gravitoelectric and gravitomagnetic time delays [15].  Moreover, $\Delta_{\Lambda}$ is the
additional time delay due to the cosmological constant.  This delay can be calculated using
the approach developed in [15]; that is, 

\begin{equation}
\Delta = \frac{1}{2c}\;\int^{P_2}_{P_1}\; h_{\alpha\beta}k^{\alpha}k^{\beta} dl\;\;,
\end{equation}

\noindent where $h_{\alpha\beta} = g_{\alpha\beta} -\eta_{\alpha\beta}\;,\;k^{\alpha}=(1,
{\bf\hat k)}, {\bf\hat k}$ is a unit vector along the direction of propagation of the ray in
the background Minkowski spactime and $dl$ is the element of straight line connecting $P_1$ to
$P_2$.  

Concentrating on the cosmological constant, we note that for $M=a=0$, equation (1) reduces to
the de Sitter metric.  Under the coordinate transformation

\begin{equation} r(\rho) = \frac{\rho}{1 + \frac{1}{12}\Lambda\rho^2}\;\;,
\end{equation}

\noindent the de Sitter metric takes the isotropic form

\begin{equation} 
ds^2 = -\left(\frac{1-\psi}{1+ \psi}\right)^2 dt^2 +
\frac{1}{(1+\psi)^2}\;(d\rho^2 + \rho^2 d\Omega^2)\;\;,
\end{equation}

\noindent where $\psi = \Lambda\rho^2/12$.  It follows from equation (16) that for the
cosmological constant alone,

\begin{equation} h_{00} = \frac{1}{3}\Lambda\rho^2\;\;,\;\;h_{0i} =
0\;\;,\;\;h_{ij}=-\frac{1}{6}\Lambda\rho^2\delta_{ij}\;\;.
\end{equation}

\noindent Using these results in equation (14), we find

\begin{equation}
\Delta_{\Lambda} = \frac{\Lambda}{12c}\;\int^{P_2}_{P_1}\rho^2 dl\;\;.
\end{equation}

\noindent It is straightforward to compute this integral and the result is 

\begin{equation}
\Delta_{\Lambda} =
\frac{\Lambda}{36c}\;|\mbox{\boldmath$\rho_1$}-\mbox{\boldmath$\rho_2$}|\;(\rho_1^2
+\mbox{\boldmath$\rho$}_1\cdot\mbox{\boldmath$\rho_2$} +
\rho^2_2)\;\;.
\end{equation}

\noindent We note that for $\Lambda\approx 10^{-56}{\rm cm}^{-2}, \Delta_{\Lambda}\sim 2$
sec for a ray of radiation crossing the disk of our Galaxy.

The calculation of first-order gravitational delays in equation (13) assumes that the ray
follows a straight line.  Therefore, there must be in addition a geometrical delay that takes
the actual deflected path into account; for weak lensing, the geometrical delay is of second
order in the deflecting potentials.  It is important to recognize that the cosmological
constant does not participate in the deflection of light rays in the Kottler spacetime [16,
17]; therefore, the cosmological constant does not contribute to the geometrical delay in the
Schwarzschild-de Sitter spacetime.  

It follows from equation (19) that locally, i.e. for distances much smaller than
$(3M/\Lambda)^{1/3}$, the time delay due to the cosmological constant is very small compared
to the corresponding Shapiro time delay.  On the other hand, equation (19) indicates that for
distances approaching $\Lambda^{-1/2}$, the time delay could be rather significant, but then
our linear approximation scheme may break down.  Thus for cosmological observations
$\Delta_{\Lambda}$ should in general be taken into account, since the present uncertainty in
the measurement of gravitational lensing time delay is about $1\over 2$ day [15].

\vspace{.25in}

\noindent 4.  Orbital precession

\vspace{.10in} The solution of the equations of motion of a test particle in the Kerr-de
Sitter spacetime is quite complicated.  On the other hand, it turns out that one can obtain
the main results regarding the gravitomagnetic clock effect, etc., from the linear
approximation of general relativity.  Therefore, we consider a linear post-Newtonian approach
using a Lagrangian of the form ${\cal L} = -m\:ds/dt$, where $m$ is the mass of the test
particle and the metric is expressed in $(t, \mbox{\boldmath$\rho$})$ coordinates, where
$\mbox{\boldmath$\rho$} = (x, y, z)$ indicates a point in space with isotropic
(post-Newtonian) Cartesian coordinates.  The equations of motion then take the form

\begin{equation}
\frac{d^2\mbox{\boldmath$\rho$}}{dt^2} + \frac{GM\mbox{\boldmath$\rho$}}{\rho^3} = {\bf
F}\;\;,
\end{equation}

\noindent where ${\bf F}$ is the post-Newtonian perturbing function.  The details of this
process of reduction of the equations of motion to the form (20) are straightforward and have
been described in detail in [4, 18].  Under certain circumstances, the orbital perturbations
can be easily characterized in terms of a modified orbit as has been done for the
gravitomagnetic clock effect [4].  To illustrate this procedure, it is instructive to digress
here and consider the case of bound motion in the Kerr-Taub-NUT spacetime; the results can
be compared and contrasted with those of the Kerr-de Sitter spacetime.  Linearizing this
metric in the angular momentum and the NUT parameter of the source, we find the Schwarzschild
metric together with the Lense-Thirring and Taub-NUT off-diagonal terms.  Introducing the
isotropic Schwarzschild radial coordinate
$\rho$ and the corresponding Cartesian coordinates, we find that this gravitoelectric
background is perturbed by a gravitomagnetic field,

\begin{equation} 
{\bf B}_g = \frac{GJ}{c\rho^5}\left[3(\mbox{\boldmath${\rho}$}\cdot{\bf\hat
J}) \mbox{\boldmath$\rho$} -\rho^2{\bf\hat J}\right]-\frac{c^2
\ell\mbox{\boldmath$\rho$}}{\rho^3}
\end{equation}

\noindent due to the angular momentum (i.e. the gravitomagnetic dipole moment J) and the
gravitomagnetic monopole moment $(-c^2 \ell)$ of the source.  Here the NUT parameter $\ell$ has
the dimension of length.  The post-Newtonian gravitomagnetic orbital perturbation can then be
obtained from equation (20), where ${\bf F}$ is of the form ${\bf F} = -2{\bf v} \times {\bf
B_g}/c$.  The linear perturbation of the orbit due to the angular momentum of the source has
been worked out in detail in [4] in connection with the gravitomagnetic clock effect. 
Therefore, we only indicate here the orbital perturbations due to ${\bf F_\ell} =
-2c\ell\mbox{\boldmath$\rho$}\times{\bf v}/\rho^3$ using the method developed in [4].  

For the motion of a test mass around a gravitomagnetic monopole, the energy of the particle
is conserved and so is its angular momentum, which is however augmented by a contribution
from the interaction of the particle with the monopole, i.e. ${\bf j} =
m\mbox{\boldmath$\rho$}\times{\bf v} - 2 mc\ell\mbox{\boldmath$\rho$}/\rho$ is the conserved
quantity.  Following the analysis given in [4], let us note that the unperturbed orbit is
given by a Keplerian ellipse in the $(X, Y)$-plane of the $(X, Y, Z)$ coordinate system that
is related to the $(x, y, z)$ system by a rotation,

\begin{eqnarray}
x &=& \cos\Omega\: X - \sin\Omega\cos i\: Y + \sin\Omega\sin i\: Z\;\;,\\
y &=& \sin\Omega\: X + \cos\Omega\cos i\: Y - \cos\Omega\sin i\: Z\;\;,\\
z &=& \sin i\: Y + \cos i\: Z\;\;,
\end{eqnarray} 

\noindent where $i$ is the inclination angle and $\Omega$ is the longitude of the line of the
ascending nodes.  The Keplerian ellipse is then given by $X=\rho\cos\varphi, Y =
\rho\sin\varphi$ and $Z=0$, where

\begin{equation}
\rho = \frac{a(1-e^2)}{1 + e\cos(\varphi - g)}\;\;\;,\;\;\;\frac{d\varphi}{dt} =
\frac{L_0}{\rho^2}
\end{equation}

\noindent and $L_0=\sqrt{GMa(1-e^2)}$ is the specific orbital angular momentum of the
unperturbed orbit.  Here $a, e$ and $g$ are respectively the semimajor axis, eccentricity and
argument of the pericenter of the unperturbed elliptical orbit.

Once the gravitomagnetic monopole moment of the source is turned on at $t=0$, when
observations begin, the only change that occurs in the orbit to linear order is that $Z$ is
no longer zero but is given by 

\begin{equation} 
Z = \frac{2}{c}\mu L_0\;\frac{1-\cos(\varphi - \varphi_0)}{1+e\cos(\varphi -
g)}\;\;,
\end{equation}

\noindent where $\mu = -c^2\ell/(GM)$ is the dimensionless strength of the monopole and
$\varphi = \varphi_0$ at $t=0$.  A detailed examination shows that there is no gravitomagnetic
clock effect in this case to first order in $\ell$.  In this linear order, the net angular
momentum ${\bf j}$ turns out to have componets $j_X = -2mc\ell\cos\varphi_0, j_Y =
-2mc\ell\sin\varphi_0$ and $j_Z = mL_0$. 

The gravitomagnetic monopole moment of the source presumably exists for all time; therefore,
the procedure described above has to be re-interpreted in terms of an {\it osculating}
ellipse.  Namely, the position and velocity of the test mass at any given instant of time
define an elliptical orbit that is momentarily tangent to the actual perturbed orbit. 
Therefore, the perturbed motion may be described in terms of the evolution of the osculating
ellipse as in the Lagrange planetary equations.  In the case under consideration here the
observations are assumed to begin at $t=0$; hence, the elliptical orbit (25) is simply the
osculating ellipse of the perturbed orbit at $t=0$.  

Let us now return to the Kerr-de Sitter metric and linearize it in the small parameters
$\epsilon = a/M$ and $\delta=\Lambda M^2$ with $M\neq 0$.  We then need to introduce an
isotropic radial coordinate $\rho$ that puts the Kottler metric in isotropic form.  The
connection between $r$ and $\rho$ is given by the differential equation

\begin{equation}
\left(\frac{dr}{d\rho}\right)^2 =
\frac{r^2}{\rho^2}\left(1-\frac{2M}{r}-\frac{\Lambda}{3}r^2\right)\;\;.
\end{equation}

\noindent To first order in $\delta$, we find that 

\begin{equation} r(\rho) = \rho\left( 1 + \frac{M}{2\rho}\right)^2 -
\frac{\delta}{24}\rho\left(1-\frac{M^2}{4\rho^2}\right) I \left(\frac{2\rho}{M}\right)\;\;,
\end{equation}

\noindent where $I(x)$ is given by 

\begin{equation} I(x) = \int^x\frac{(1+ u)^6}{(1-u)^2}\;\frac{du}{u^3}\;\;.
\end{equation}

\noindent This integral can be evaluated and the result is 

\begin{equation} 
2I(x) = K + x^2 + 16x + 60\:{\rm ln}\:x - \frac{128}{x-1} -
\frac{16}{x}-\frac{1}{x^2}\;\;,
\end{equation}

\noindent where $K$ is a dimensionless integration constant that we can set equal to zero for
the sake of simplicity.  A detailed examination of the resulting equation of motion then
shows that the dominant effect of the cosmological constant would simply be due to the
effective Newtonian perturbation of the form ${\bf F} = \lambda\mbox{\boldmath$\rho$}$, where
$\lambda=\Lambda c^2/3$.  Under such a perturbation the orbital angular momentum is conserved
and the orbit is thus planar.  On the other hand, the orbital energy  is augmented by a
contribution from the cosmological constant so that the conserved quantity is $E$, where $2E =
v^2 - \lambda\rho^2 - 2GM/\rho$.  Though the perturbing function is simply proportional to
$\mbox{\boldmath$\rho$}$ in this  case, it turns out that the simple method developed in [4] is
not directly applicable in this case for orbits of arbitrary eccentricity, since the
perturbing function has the form of an infinite series in powers of the eccentricity (cf.
section 5).  This is in contrast with the Kerr-Taub-NUT case.  It is therefore necessary to
adopt a different approach to the motion of the perturbed orbit unless $e<<1$, in which case
the method of [4] is adequate.  The general solution to first order in the cosmological
constant is considered in the next section; in the following, we study the average motion of
the perturbed orbit.

It is interesting to compute the average rate of precession of the pericenter under the
influence of $\Lambda$.  To this end, let us introduce the Runge-Lenz vector of the perturbed
orbit  

\begin{equation} 
{\bf Q} = {\bf v} \times {\bf L} - GM\mbox{\boldmath$\rho$}/\rho\;\;,
\end{equation}

\noindent where ${\bf L} = L_0{\bf\hat Z}$ and we note that for ${\bf F}=0, {\bf Q_0} =
GMe(\cos g, \sin g, 0)$ in the $(X, Y, Z)$ coordinate system.  For ${\bf F} =
\lambda\mbox{\boldmath$\rho$}$, the orbit remains in the $(X, Y)$-plane and ${\dot Q}_X =
\lambda L_0 Y, {\dot Q}_Y = -\lambda L_0X$ and ${\dot Q}_Z = 0$.  Thus in this case

\begin{equation}
\frac{d{\bf Q}}{dt} = -\lambda {\bf L}\times\mbox{\boldmath$\rho$}\;\;.
\end{equation}

\noindent Averaging this relation over the ``fast'' motion, i.e. the unperturbed orbital
motion of the test particle (with period $T$) would reveal the ``slow'' (secular) behavior of
the Runge-Lenz vector over a long period of time.  Defining the average over a period as 

\begin{equation} <f> = \frac{1}{T}\int^T_0 fdt = (1-e^2)^{3/2}
\;\frac{1}{2\pi}\int^{2\pi}_0\frac{f(\varphi)d\varphi}{[1+e\cos(\varphi-g)]^2}\;\;,
\end{equation}

\noindent we obtain

\begin{equation} <X> = -\frac{3}{2}ae\cos g\;\;,\;\;<Y> = -\frac{3}{2}ae\sin g\;\;,
\end{equation}

\noindent where we have used the relation 

\begin{equation}
\frac{1}{2\pi}\int^{\zeta_0+2\pi}_{\zeta_0}\frac{\cos\zeta d\zeta}{(1+ e\cos\zeta)^3} =
-\frac{3}{2}\;\frac{e}{(1-e^2)^{5/2}}\;\;.
\end{equation}

\noindent Thus we find that 

\begin{equation} 
<\frac{d{\bf Q}}{dt}> = \mbox{\boldmath$\Omega$}\times{\bf Q_0}\;\;,
\end{equation}

\noindent where

\begin{equation}
\mbox{\boldmath $\Omega$} = \frac{3}{2}\;\frac{a\lambda L_0}{GM}{\bf\hat Z} =
\frac{1}{2}\;\frac{\Lambda c^2}{\omega}\sqrt{1-e^2}{\bf\hat Z}\;\;.
\end{equation}

\noindent Here $\omega = 2\pi/T$ is the Keplerian frequency of the unperturbed orbit.  For
$\Lambda\approx 10^{-56} {\rm cm}^{-2}$, the precession frequency (37) is too small to be
observable at present in the solar system [16].  The precession angle after one period is given
by 

\begin{equation}
\frac{\pi c^2\Lambda a^3}{GM}\sqrt{1-e^2}\;\;.
\end{equation}

\noindent To first order in eccentricity, this result agrees with previous work [16, 19],
while the exact dependence on eccentricity as $\sqrt{1-e^2}$ differs from
$(1-e^2)^3$ given in [19].  

To gain a deeper understanding of the average behavior of the orbit, we consider the
osculating ellipse in this case and note that the orbital elements of this ellipse vary with
time according to the Lagrange planetary equations [20] as 

\begin{eqnarray}
\frac{da}{dt}&=&\frac{2\lambda ae\sqrt{1-e^2}}{\omega}\;\frac{\sin(\varphi -
g)}{1+e\cos(\varphi - g)}\;\;,\\
\frac{de}{dt}&=&\frac{1-e^2}{2ea}\;\frac{da}{dt}\;\;,\\
\frac{dg}{dt}
&=& -\frac{\lambda\sqrt{(1-e^2)^3}}{e\omega}\;\frac{\cos(\varphi-g)}{1+e\cos(\varphi-g)}\;\;.
\end{eqnarray}

\noindent Employing the averaging method once again, we find that 

\begin{equation} <\frac{da}{dt}> = 0\;\;,\;\; <\frac{de}{dt}> = 0\;\;,
\end{equation}

\begin{equation} <\frac{dg}{dt}> = \frac{3}{2}\;\frac{\lambda}{\omega}\sqrt{1-e^2}\;\;,
\end{equation}

\noindent where relation (35) has been used.  The rate of advance of the pericenter given
by equation (43) is consistent with our previous result.  Thus in the presence of a
cosmological constant the orbital plane does not change and the orbit keeps its shape on the
average while precessing with frequency $\Lambda c^2\sqrt{1-e^2}/(2\omega)$.

It follows from equation (37) that the frequency of the pericenter precession caused by the
cosmological constant is directly proportional to the period of the binary orbit, which must
therefore be very large for the effect to be perceptible.  The ratio of this precession
frequency to the Einstein pericenter precession frequency is given by $\Lambda
a^{4}c^{4}(1-e^{2})^{3/2}/(6G^2M^2)$, which is $\sim 10^{-14}$ for the motion of the Earth
around the Sun.  We conclude that the influence of the cosmological constant on orbital motion
is too small to be measurable in the solar system. 

\vspace{.25in}

\noindent 5.  Perturbed Keplerian motion

\vspace{.10in} 
It is interesting to give the general solution of equation (20) for ${\bf F} =
\lambda\mbox{\boldmath $\rho$}$ to first order in the cosmological constant.  For a
positive cosmological constant, the effective force on the test mass is along the radial
direction away from the source.  The orbit is thus planar and
$\rho^2 d\varphi/dt=L_0$ is the constant specific orbital angular momentum.  In terms of
$u=\rho^{-1}$, the radial part of equation (20) reduces in this case to 

\begin{equation}
\frac{d^2u}{d\varphi^2} + u = p^{-1} - \frac{\lambda}{L^2_0 u^3}\;\;,
\end{equation}

\noindent where $p = L^2_0/(GM)$ is a positive constant.  The right side of equation (44) can
be expressed as a function of $\varphi$ if we substitute for $u$ the unperturbed solution in
accordance with our linear perturbation scheme.  The unperturbed solution is given by 

\begin{equation} 
u_0 = \frac{1}{p}\left(1+e\cos\varphi\right)\;\;,
\end{equation}

\noindent where we have chosen the planar Cartesian coordinates such that $g=0$ for the sake
of simplicity.  Here $e\geq0$ is the orbital eccentricity of the conic section.  For an
ellipse $e<1$ and $p=a(1-e^2)$, where the pericenter distance from the focus is
$\rho_{min}=a(1-e)$ and $a$ is the semimajor axis.  For a parabola $e=1$ and $\rho_{min} =
p/2$ is the pericenter distance, while for a hyperbola $e>1$ and $p=a(e^2-1)$, where the
pericenter distance from the focus is $\rho_{min} = a(e-1)$.  Substituting equation (45) for
$u$ in the right side of equation (44), the resulting equation can be transformed to the form 

\begin{equation} (1-\xi^2)\;\frac{d^2U}{d\xi^2} - \xi \frac{dU}{d\xi} + U =
\frac{q}{(1+e\xi)^3}\;\;,
\end{equation}

\noindent where $\xi=\cos\varphi, U = u-p^{-1}$ and $q = -\lambda p^3/L^2_0$.  Equation (46)
can be integrated once and we find

\begin{equation} 
(1-\xi^2)\frac{dU}{d\xi} + \xi U = -\frac{q}{2e(1+e\xi)^2} + C\;\;,
\end{equation}

\noindent where $C$ is an integration constant.  Equation (47) can be written in the form

\begin{equation}
\frac{d}{d\xi}\left[(1-\xi^2)^{-1/2} U\right] = -\frac{q}{2e(1-\xi^2)^{3/2}(1+e\xi)^2} +
\frac{C}{(1-\xi^2)^{3/2}}
\end{equation}

\noindent and integrated using formulas 2.264, 2.266 and 2.268 given in [21].  The result for
an {\it elliptical} orbit is 

\begin{equation}
U = \frac{1}{2}\frac{q}{(1-e^2)^2}\left[3(1+ e {\cal I}\sin\varphi) -
\frac{1-e^2}{1+e\cos\varphi} - (2e+ \frac{1}{e})\cos\varphi\right]+ C\cos\varphi +
S\sin\varphi\;\;,
\end{equation}

\noindent where $S$ is a constant of integration and 

\begin{equation}
{\cal I} = \frac{1}{\sqrt{1-e^2}} \;{\rm
arc}\sin\left(\frac{e+\cos\varphi}{1+e\cos\varphi}\right)\;\;.
\end{equation}

\noindent The integration of equation (48) can be similarly carried out for parabolic and
hyperbolic orbits using the relevant formulas of [20].  We expect that $u$ reduces to the
unperturbed solution $u_0$ as $\lambda\rightarrow 0$; therefore, it is useful to set
$C=qC^{\prime} + e/p$ and $S=qS^{\prime}$.  The constants $C^{\prime}$ and $S^{\prime}$ can be
uniquely determined using the initial conditions, i.e. the initial position and velocity of
the test mass.  In this way, we find the orbital radius $\rho = u^{-1} = (p^{-1} + U)^{-1}$
and then
$\varphi(t)$ by integrating $d\varphi/dt=L_0\rho^{-2}$. 

\vspace{.25in}

\noindent 6.  Discussion

\vspace{.10in}

In this paper we have considered some of the local physical consequences of the existence of a
cosmological constant.  In particular, we have considered the motion of test bodies and the
propagation of rays in the Kerr-de Sitter spacetime.  While the clock effect is rather weakly
affected by the presence of a cosmological constant, there is a (gravitoelectric) time delay
in the propagation of rays directly proportional to $\Lambda$.  Though this effect is
too small to be measurable in the solar system, it should in general be taken into
account in the interpretation of gravitational lensing time delay.  Moreover, the average
behavior of bounded orbits has been examined and the general rate of orbital precession has
been determined for an arbitrary eccentric orbit.  On the average, the orbital plane remains
invariant and the orbit keeps its shape but precesses due to the presence of a cosmological
constant; however, the precession rate is too small to be detectable in the solar system. 

\vspace{.25in}

\noindent References

\vspace{.10in}
\begin{description}

\item{[1]} Cohen JM and Mashhoon B 1993 {\it Phys. Lett. A} {\bf 181} 353
\item{[2]} Mashhoon B, Gronwald F and Theiss DS 1999 {\it Ann. Phys. (Leipzig)} {\bf 8} 135
\item{[3]} Mashhoon B, Gronwald F and Lichtenegger HIM 2000 in:  {\it Gyros, Clocks,
Interferometers...:  Testing Relativistic Gravity in Space}, edited by C. L\"{a}mmerzahl,
C.W.F. Everitt and F.W. Hehl (Berlin : Springer), pp. 83-108
\item{[4]} Mashhoon B, Iorio L and Lichtenegger H 2001 {\it Phys. Lett. A} {\bf 292} 49
\item{[5]} Demia\'{n}ski M 1973 {\it Acta Astron.} {\bf 23} 197
\item{[6]} Carter B 1973 in:  {\it Black Holes}, edited by C. DeWitt and B.S. DeWitt (New
York:  Gordon and Breach), pp. 101-103
\item{[7]} Kottler F 1918 {\it Ann. Phys. (Leipzig)} {\bf 56} 410
\item{[8]} Schmidt H-J 1993 {\it Fortschr. Phys.} {\bf 41} 179
\item{[9]} Liu H and Mashhoon B 1995 {\it Ann. Phys. (Leipzig)} {\bf 4} 565
\item{[10]} Peebles PJE and Ratra B 2003 {\it Rev. Mod. Phys.}, in press; astro-ph/0207347
\item{[11]} Howes RJ 1981 {\it Gen. Rel. Grav.} {\bf 13} 829
\item{[12]} Bini D, Jantzen RT and Mashhoon B 2001 {\it Class. Quantum Grav.} {\bf 18} 653
\item{[13]} Bini D, Jantzen RT and Mashhoon B 2002 {\it Class. Quantum Grav.} {\bf 19} 17
\item{[14]} Maartens R, Mashhoon B and Matravers DR 2002 {\it Class. Quantum Grav.} {\bf 19}
195
\item{[15]} Ciufolini I, Kopeikin S, Mashhoon B and Ricci F 2003 {\it Phys. Lett. A} {\bf
308} 101
\item{[16]} Islam JN 1983 {\it Phys. Lett. A} {\bf 97} 239
\item{[17]} Lake K 2002 {\it Phys. Rev. D} {\bf 65} 087301
\item{[18]} Mashhoon B 2001 in:  {\it Reference Frames and Gravitomagnetism}, edited by J.-F.
Pascual-S\'{a}nchez, L. Flor\'{i}a, A. San Miguel and F. Vicente (Singapore: World
Scientific), pp. 121-132
\item{[19]} Rindler W 2001 {\it Relativity:  Special, General, and Cosmological} (Oxford: 
Oxford University Press), p. 305
\item{[20]} Danby JMA 1988 {\it Fundamentals of Celestial Mechanics}, 2nd ed. (Richmond: 
Willmann-Bell)
\item{[21]} Gradshteyn IS and Rhyzhik M 1980 {\it Table of Integrals, Series and Products}
(New York:  Academic Press)

\end{description}

\end{document}